\documentstyle[seceq]{ptptex}
\newcommand{\dslash}{\ooalign{\hfil/\hfil\crcr$\partial$}}
\makeatletter
\def\eqnarray{%
 \stepcounter{equation}%
 \let\@currentlabel=\theequation
 \global\@eqnswtrue
 \global\@eqcnt\z@
 \tabskip\@centering
 \let\\=\@eqncr
 $$\halign to \displaywidth\bgroup\@eqnsel\hskip\@centering
 $\displaystyle\tabskip\z@{##}$&\global\@eqcnt\@ne
 \hfil$\displaystyle{{}##{}}$\hfil
 &\global\@eqcnt\tw@$\displaystyle\tabskip\z@{##}$\hfil
 \tabskip\@centering&\llap{##}\tabskip\z@\cr}
\makeatother
\title{Bosonic Structure of a 2-Dimensional Fermion Model\\
with Interaction among Different Speices}

\author{Jiro {\sc Sakamoto}\thanks{E-mail
address:\ jsakamot@riko.shimane-u.ac.jp}\ \ and Yasunari {\sc Heike}}
\inst{Physics department, Shimane University, Matsue 690-8504, Japan}
\abst{
 We study a massive Thirring-like model in 2-dimensional space-time,
 which contains two different species of fermions. This model
is a field theoretical version of the quantum mechanical model originally 
proposed by Gl\"{o}ckle, Nogami and Fukui, where different fermions
interact with each other through $\delta$-function potentials. We derive
 a corresponding boson model by the bosonization technique in the path integral
 formulation.  This is a simple  but non-trivial extension of the freedom of
 the bosonization technique. Operator correspondences between fermion and
 boson fields are given. One of these could not be realistically expected from 
 the naive correspondence of the original single-species models. It is
 essential for this point that in our model fermions of the same kind 
 do not interact with each other directly. We find that for a specific
 value of the coupling constant, one boson field becomes free while the other 
 is a Sine-Gordon field. For this case, therefore, our two-species model
 is equivalent to the ordinary Sine-Gordon model of a single boson field. 
}

\begin{document}
\maketitle
\section{Introduction}
In previous papers we have studied relativistic bound states of a
1-space  quantum mechanical system  containing different species of 
massive fermions in order to investigate the relativistic effects for such
composite systems.\cite{sakamoto} This model is an extended one from the model of two kinds
of fermions originally proposed by Gl\"{o}ckle, Nogami and Fukui (GNF).\cite{GNF}
The Hamiltonian of this model is given by
\begin{equation}
 H=\sum_{i=1}^{n}\{-i\alpha_i P_i + m\beta_i\}
-\frac{g}{2}\sum_{i\neq j}^n(1-\alpha_i\alpha_j)\delta (x_i-x_j),
\label{hamiltonian}
\end{equation}
where $i$ and $j$ denote fermion species. 
It is essential in this model that different fermions interact  with
each other through the $\delta$-function  potentials, while fermions of the
same kind do not interact with each other directly. All the  requirements
of quantum mechanics and special relativity are satisfied.  We found an
exact solution  for $n$-body bound state which contains $n$ different
particles.

The GNF model is, however, based on the single-electron theory, where
anti-particles are not supposed to exist, and necessarily its Hamiltonian
(\ref{hamiltonian}) is not positive-definite. One way to overcome this 
defect would be to go into field theory. It is seen that the GNF model
can be derived from a massive Thirring-like model  in 2-dimensional
space-time, which we will give in the next section. 

One of the powerful approaches to study a 2-dimensional fermion system is
to utilize the bosonization technique,\cite{bosonization} though it may be
applicable only for  charge-zero sectors of fermions, i.e for 
sectors of particle and anti-particle pairs.
With this technique, one can expose hidden properties of such a fermion 
system by deriving the equivalent boson system. Since Coleman's
discovery of the equivalence between the  massive Thirring and the 
Sine-Gordon models,\cite{Coleman} many people have tried to  expand the
freedom by introducing a non-Abelian group,  by going into higher dimensional
space-time, and so on. Formulations in  most of these cases, however, 
turn out to be very complicated.\cite{bosonization} 

In this article we  apply the  bosonization technique to the field
theoretical GNF model with  path integral quantization in order to
investigate the physical structures of this model. As  will be shown in the
following, this is one of easier, but non-trivial, directions to expand
the freedom of the system in the bosonization technique. We will study
here the case of two species of fermions as the simplest case. Extension
to  the case of more than two species will be discussed elsewhere.
We find in this model that for a specific value of the coupling
constant one of the boson fields becomes free, while the other satisfies
the Sine-Gordon equation. For that value of coupling constant, therefore, 
our two-species model is equivalent to the ordinary Sine-Gordon model of
a single species.
  
We use the metric convention in Minkowski space-time, $g_{\mu\nu}=(-1,+1)$ and
$\epsilon^{01}=-\epsilon_{01}=1$. Gamma-matrices are given as $\gamma^0
 = -\gamma_0=i\sigma_x$, $\gamma^1 = \gamma_1=\sigma_y$,
$\gamma_5 =\gamma_0\gamma_1 =\sigma_z$, where $\sigma_x, \sigma_y$ and 
$\sigma_z$ are the Pauli matrices.

\setcounter{equation}{0}
\section{Model}
Our initial Lagrangian is given by
\begin{equation}
 {\cal L}= 
\overline{\psi}_a({\ooalign{\hfil/\hfil\crcr$\partial$}}-m)\psi_a+\overline{\psi}_b({\ooalign{\hfil/\hfil\crcr$\partial$}}-m)\psi_b
+ {1\over 2}gj_{a\mu}j_b^{\mu},\label{lagrangian1}
\end{equation}
where $a$ and $b$ denote the fermion species and 
\begin{equation}
 j_{a(b)\mu }= i\overline{\psi}_{a(b)}\gamma_{\mu}\psi_{a(b)}.\label{current_a}
\end{equation}
In (\ref{lagrangian1}), neither $a$- nor $b$-fermion interacts with
itself directly. As mentioned in the previous section, one can
derive the quantum mechanical GNF model from this Lagrangian. If we
include a self-coupling term like $(\sum_{i=a,b}j_{i\mu})^2$ instead of
$j_{a\mu}j_b^\mu$ in (\ref{lagrangian1}) above, which is one
of the diagonal terms of $SU(2)$, the model becomes simple, because we need
only a single auxiliary vector field for such a case, unlike
(\ref{lagrangian2}) below. We can suppose $g\geq 0$ without loss of generality
because we can change its sign by charge conjugation of one of fermion
species, $a$ or $b$.
 
We introduce auxiliary vector fields $A_{\mu}$ and $B_{\mu}$ to rewrite
the Lagrangian (\ref{lagrangian1}) as
\begin{equation}
 {\cal L}= 
\overline{\psi}_a({\ooalign{\hfil/\hfil\crcr$\partial$}}-m)\psi_a+\overline{\psi}_b({\ooalign{\hfil/\hfil\crcr$\partial$}}-m)\psi_b
 +\frac{1}{2}gA_{\mu}j_a^{\mu} +\frac{1}{2}gB_{\mu} j_b^{\mu} -\frac{1}{2}gA_{\mu}B^{\mu}. \label{lagrangian2}
\end{equation}
In 2-dimensional space-time, we can write  vector fields $A_{\mu}$ and
$B_{\mu}$ with scalar fields $\phi$ and $\chi$ as
\begin{eqnarray}
A_{\mu}&=&\epsilon_{\mu\nu}\partial^{\nu}\phi_a + \partial_{\mu}\chi_a,\nonumber \\
B_{\mu}&=&\epsilon_{\mu\nu}\partial^{\nu}\phi_b + \partial_{\mu}\chi_b.
\end{eqnarray}
Then we have
\begin{eqnarray}
 {\cal L}&=&\overline{\psi}_a({\ooalign{\hfil/\hfil\crcr$\partial$}}+\frac{ig}{2}
\epsilon_{\mu\nu}\partial^{\nu}\phi_a\gamma^\mu +
\frac{ig}{2}{\ooalign{\hfil/\hfil\crcr$\partial$}}\chi_a - m)\psi_a 
+(a\rightarrow b) \nonumber\\
& & +\frac{g}{2}(\partial^{\mu}\phi_a\partial_{\mu}\phi_b 
- \partial_{\mu}\chi_a\partial^{\mu}\chi_b).  \label{lagrangian3}
\end{eqnarray}
We transform the fermion variables as 
\begin{eqnarray}
 \psi_a & \rightarrow & \psi_a '
=\exp\left\{\frac{ig}{2}(-\gamma_5\phi_a+\chi_a)\right\}\psi_a, \nonumber\\
\psi_b & \rightarrow & \psi_b '
=\exp\left\{\frac{ig}{2}(-\gamma_5\phi_b+\chi_b)\right\}\psi_b,\label{chiral_trans}
\end{eqnarray}
to rewrite the Lagrangian as
\begin{eqnarray}
 {\cal L}' &=& \overline{\psi}_a'({\ooalign{\hfil/\hfil\crcr$\partial$}}-m
\exp\{ig\gamma_5\phi_a\})\psi_a' + 
 \overline{\psi}_b'({\ooalign{\hfil/\hfil\crcr$\partial$}}-m
\exp\{ig\gamma_5\phi_b\})\psi_b'  
 \nonumber \\
&  & +\frac{g}{2}(\partial^{\mu}\phi_a\partial_{\mu}\phi_b -
 \partial^{\mu}\chi_a\partial_{\mu}\chi_b).
\end{eqnarray}
Though transformation (\ref{chiral_trans}) causes the $\chi$ to decouple
from the other fields,  it changes the path-integral measure of the fermion
fields as
\begin{equation}
 \prod d\overline{\psi}_a d\psi_a = \prod  d\overline{\psi}_a'd\psi_a'\det\mid
\exp\{-ig\gamma_5\phi_a\}\mid,
\end{equation}
and following Fujikawa's procedure\cite{Fujikawa} this determinant is calculated as
\begin{equation}
 \det\mid\exp\{-ig\gamma_5\phi_a\}\mid = \exp\left\{i\int d^2 x 
\frac{g^2}{4\pi}(\partial\phi_a)^2\right\}.
\end{equation}
Then, the effective Lagrangian is given by
\begin{eqnarray}
  {\cal L_{\rm eff}} &=& \overline{\psi}_a({\ooalign{\hfil/\hfil\crcr$\partial$}}-m
\exp\{ig\gamma_5\phi_a\})\psi_a + 
 \overline{\psi}_b({\ooalign{\hfil/\hfil\crcr$\partial$}}-m
\exp\{ig\gamma_5\phi_b\})\psi_b  
 \nonumber \\
&  & +\frac{g}{2}\partial^{\mu}\phi_a\partial_{\mu}\phi_b 
 + \frac{g^2}{4\pi}\{(\partial\phi_a)^2 + (\partial\phi_b)^2\},\label{L_eff}
\end{eqnarray}
and the generating functional $Z$ of the Green function is 
\begin{equation}
 Z = \int \prod_{i=a,b}d\overline{\psi}_i d\psi_i d\phi_i
\exp i\int d^2x{\cal L_{\rm eff}}.\label{Z}
\end{equation}
Here and hereafter we write  $\psi$ for $\psi '$. 
By the `Wick' rotation we transform ourselves into  Euclidean space-time
from the Minkowski space-time. Then the generating functional (\ref{Z}) is 
rewritten 
as
\begin{eqnarray}
 Z_E &=& \int \prod_{i=a,b}d\overline{\psi}_i d\psi_i d\phi_i
\exp\left( -\int d^2x \Bigl[ \overline{\psi}_a({\ooalign{\hfil/\hfil\crcr$\partial$}}-
me^{ig\gamma_5\phi_a})\psi_a + (a \rightarrow b) \right.\nonumber\\
 & &\left. +\frac{g}{2}\partial_{\mu}\phi_a\partial_{\mu}\phi_b
 + \frac{g^2}{4\pi}\{(\partial\phi_a)^2 + (\partial\phi_b)^2\}\Bigr]\right).\label{Z_E}
\end{eqnarray}

\section{Perturbative expansion}
Before we expand $Z_E$ of (\ref{Z_E}) in power of $m$, we transform
$(\phi_a, \phi_b)$ to $(\phi_a', \phi_b')$ as
\begin{eqnarray}
 \phi_a & = & \frac{\sqrt{2}}{2}(\phi_a'+\phi_b'), \nonumber \\
\phi_b & = & \frac{\sqrt{2}}{2}(\phi_a'-\phi_b')
\end{eqnarray}
to rewrite the Lagrangian (\ref{L_eff}) as
\begin{eqnarray}
{\cal L_{\rm eff}} &=& \overline{\psi}_a({\ooalign{\hfil/\hfil\crcr$\partial$}}-me^{ig\gamma_5\phi_a})\psi_a +
 \overline{\psi}_b({\ooalign{\hfil/\hfil\crcr$\partial$}}-m
e^{ig\gamma_5\phi_b})\psi_b
 \nonumber \\
&  & +\frac{g}{4}\left(1+\frac{g}{\pi}\right)(\partial\phi_a')^2
 + \frac{g}{4}\left(1-\frac{g}{\pi}\right)(\partial\phi_b')^2. \label{L_eff2}
\end{eqnarray}
Now, we make a perturbative expansion of $Z_E$ in (\ref{Z_E}) in $m$:
\begin{eqnarray}
Z_E &=& \int\prod d\overline{\psi}d\psi d\phi ' \exp\Bigl( -\int d^2 x {\cal L}_{\rm eff}(m=0)\Bigr) \nonumber\\
&\times& \sum_{n=0}^{\infty}\frac{m^n}{n!}\left\{\int d^2x\left(\overline{\psi}_ae^{ig\gamma_5(\phi_a'+\phi_b')\frac{\sqrt{2}}{2}}\psi_a +\overline{\psi}_be^{ig\gamma_5(\phi_a'-\phi_b')\frac{\sqrt{2}}{2}}\psi_b\right)\right\}^n.
\end{eqnarray} 
To obtain a boson model corresponding to our fermion model, we begin by
calculating  the first few terms of the expansion. It is seen
that the $n=1$  term vanishes due to the traceless property of the
$\gamma$-matrices. The $n=2$ term is  given by
\begin{eqnarray}
 Z_E^{(2)}&=& \int\prod d\overline{\psi}d\psi d\phi ' \exp\Bigl( -\int d^2 x {\cal L}_{\rm eff}(m=0)\Bigr) \nonumber\\
&&\times \frac{m^2}{2}\left\{\int d^2x\left(\overline{\psi}_ae^{ig\gamma_5(\phi_a'+\phi_b')\frac{\sqrt{2}}{2}}\psi_a +\overline{\psi}_b e^{ig\gamma_5(\phi_a'-\phi_b')\frac{\sqrt{2}}{2}}\psi_b\right)\right\}^2 \nonumber\\
&=& -\frac{m^2}{2}\left\{\left\langle {\rm Tr}\left(\dslash^{-1}e^{ig\gamma_5(\phi_a'+\phi_b')\frac{\sqrt{2}}{2}}\cdot \dslash^{-1}e^{ig\gamma_5(\phi_a'+\phi_b')\frac{\sqrt{2}}{2}}\right)\right\rangle \right. \nonumber\\
&&+\left\langle{\rm Tr}\left(\dslash^{-1}e^{ig\gamma_5(\phi_a'-\phi_b')\frac{\sqrt{2}}{2}}\cdot\dslash^{-1}e^{ig\gamma_5(\phi_a'-\phi_b')\frac{\sqrt{2}}{2}}\right)\right\rangle \nonumber\\
&& \left. -2\left\langle\left({\rm Tr}\dslash^{-1}e^{ig\gamma_5(\phi_a'+\phi_b')\frac{\sqrt{2}}{2}}\right)\left({\rm Tr}\dslash^{-1}e^{ig\gamma_5(\phi_a'-\phi_b')\frac{\sqrt{2}}{2}}\right)\right\rangle \right\}.\label{z_e2}
\end{eqnarray}
Here $\langle\cdots\rangle$ denotes the vacuum expectation value of the
time-ordered product, and Tr represents the  trace with respect to space-time
coordinates and $\gamma$-matrices. We have made use of the fact that the 
fermion propagator is given by
\begin{equation}
 \langle\overline{\psi}(x)\psi (y)\rangle = \dslash^{-1}(x,y)= \frac{1}{2\pi}\frac{\gamma\cdot (x-y)}{(x-y)^2}.
\end{equation}
It is easily seen that the third term of Eq.~(\ref{z_e2}) vanishes due to
the traceless property of $\gamma$-matrices, and that the first and second
terms are equal.  The first (or second) term of Eq.~(\ref{z_e2}) can be
rewritten as
\begin{eqnarray}
&& \frac{m^2}{8\pi^2}\int d^2xd^2y\frac{1}{(x-y)^2} {\rm tr}\left\langle e^{ig\gamma_5\frac{\sqrt{2}}{2}\{\phi_a'(x)\pm\phi_b'(x)-\phi_a'(y)\mp\phi_b'(y)\}}
\right\rangle \nonumber\\
&&=\frac{m^2}{8\pi^2}\int d^2xd^2y\frac{1}{(x-y)^2}\nonumber\\
&&\times\exp {\frac{g^2}{2}\{\langle\phi_a'(x)\phi_a'(y)\rangle-\langle{\phi_a'} ^2 \rangle+\langle\phi_b'(x)\phi_b'(y)\rangle-\langle{\phi_b'}^2\rangle\}}.
\end{eqnarray}
Here tr represents the $\gamma$-matrix trace. From Eq.~(\ref{L_eff2}) the
boson propagators are given as
\begin{eqnarray}
\langle\phi_a'(x)\phi_a'(y)\rangle &\equiv&\Delta_a'(x-y)= -\lambda_+\ln(x-y)^2\mu^2,\nonumber\\
\langle\phi_b'(x)\phi_b'(y)\rangle &\equiv&\Delta_b'(x-y)= -\lambda_-\ln(x-y)^2\mu^2,
\end{eqnarray}
where $\mu$ is a small infrared cut-off mass, which will be set  to zero
after the calculations, and we have put
\begin{equation}
 \lambda_{\pm}=\frac{1}{2g(g\pm\pi)}.
\end{equation}
Making use of the identity
\begin{eqnarray}
 \frac{1}{(x-y)^2}=\mu^2\exp\frac{1}{2}\left\{\frac{1}{\lambda_+}\Delta_a'(x-y)+\frac{1}{\lambda_-}\Delta_b'(x-y)\right\}, \label{conv_prop}
\end{eqnarray}
we have
\begin{eqnarray}
 Z_E^{(2)}&=&\frac{m^2\mu^2}{2\pi^2}e^{\{(g^2+g\pi)\langle{\phi_a'}^2\rangle+
(g^2-g\pi)\langle{\phi_b'}^2\rangle\}}\nonumber\\
&&\times \int d^2xd^2y\exp \left(\frac{3}{2}g^2+g\pi\right)\{\langle\phi_a'(x)\phi_a'(y)\rangle-\langle{\phi_a'}^2\rangle\}\nonumber\\
&&\times \exp\left(\frac{3}{2}g^2-g\pi\right)\{\langle\phi_b'(x)\phi_b'(y)\rangle-\langle{\phi_b'}^2\rangle\}\nonumber\\
&=&\frac{m^2\mu^2z}{2\pi^2}\int d^2xd^2y \left\langle\cos\sqrt{3g^2/2+g\pi}\{\phi_a'(x)-\phi_a'(y)\}\right\rangle\nonumber\\
&&\times \left\langle\cos\sqrt{3g^2/2-g\pi}\{\phi_b'(x)-\phi_b'(y)\}\right\rangle,\label{Z_E3}
\end{eqnarray}
where we put\footnote{This term is, of course, divergent. We suppose that
boson propagators are properly regularized to remove ultra-violet divergences.}
\begin{equation}
 z=\exp \{(g^2+g\pi)\langle{\phi_a'}^2\rangle+(g^2-g\pi)\langle{\phi_b'}^2\rangle\}.
\end{equation}
Setting ${m'}^4=4m^2\mu^2z/\pi^2$, we can rewrite Eq.~(\ref{Z_E3}) as
\begin{equation}
 Z_E^{(2)}=\frac{1}{2}\left\langle\left\{{m'}^2\int d^2x\cos\sqrt{3g^2/2+g\pi}\phi_a'\cos\sqrt{3g^2/2-g\pi}\phi_b'\right\}^2\right\rangle.
\end{equation}
This term is equal to the ${m'}^4$-order term of the perturbative
expansion of the boson model given by
\begin{eqnarray}
 {\cal L}&=&\frac{g}{4}\left(\frac{g}{\pi}+1\right)(\partial\phi_a')^2 + \frac{g}{4}\left(\frac{g}{\pi}-1\right)(\partial\phi_b')^2 \nonumber\\
&&+ {m'}^2\cos\left\{\sqrt{3g^2/2+g\pi}\phi_a'\right\}\cos\left\{\sqrt{3g^2/2-g\pi}\phi_b'\right\}.\label{bose_lag}
\end{eqnarray}
One should note that the ${m'}^2$-order term in the expansion of the above
model vanishes because of the super selection rule for the massless boson field in
 two-dimensional space-time, i.e. $\langle\exp\sum\beta\phi\rangle$
vanishes unless $\sum\beta=0$.\cite{Coleman,bosonization} In terms of
the original boson fields, $\phi_a$ and $\phi_b$, the Lagrangian above
can be rewritten as
\begin{eqnarray}
&{\cal L}&= \frac{g^2}{4\pi}(\partial\phi_a)^2 +\frac{g^2}{4\pi}(\partial\phi_b)^2 + \frac{g}{2}\partial^\mu\phi_a\partial_\mu\phi_b \nonumber\\
&+&{m'}^2\cos\left\{\sqrt{\frac{3g^2}{4}+\frac{g\pi}{2}}(\phi_a+\phi_b)\right\}\cos\left\{\sqrt{\frac{3g^2}{4}-\frac{g\pi}{2}}(\phi_a-\phi_b)\right\}. \label{bose_lag_orig}
\end{eqnarray}

In the Appendix we show that  $Z_E$ of (\ref{Z_E}) is
equal to that from (\ref{bose_lag}) order by order in the perturbative
expansion of $m$ or $m'$. We conclude, therefore, that the fermion 
model of Eq.~(\ref{lagrangian1}) is equivalent to the boson model of
(\ref{bose_lag}) or (\ref{bose_lag_orig}).

\setcounter{equation}{0}
\section{Discussion}
From the line of reasoning of the previous section, one might conclude
that one can find the corresponding boson model by calculating only the
first non-vanishing term of the perturbative expansion. But this is not the
case for multi-species models like here. In fact, instead of
(\ref{conv_prop}), with the parameter $\alpha$ we can write
\begin{equation}
 \frac{1}{(x-y)^2}=\mu^2\exp\left\{\frac{\alpha}{\lambda_+}\Delta_a'(x-y)+\frac{1-\alpha}{\lambda_-}\Delta_b'(x-y)\right\},
\end{equation} 
where $\alpha = 1/2$ for (\ref{conv_prop}). Then the
Lagrangian of the corresponding boson model for this case seems to be given by
\begin{eqnarray}
{\cal L}&=&\frac{g}{4}\left(\frac{g}{\pi}+1\right)(\partial\phi_a')^2 + \frac{g}{4}\left(\frac{
g}{\pi}-1\right)(\partial\phi_b')^2 \nonumber\\
&&+{m''}^2\cos\left\{\phi_a'\sqrt{\frac{\alpha}{\lambda_+}+\frac{g^2}{2}}\right\}
\cos\left\{\phi_b'\sqrt{\frac{1-\alpha}{\lambda_-}+\frac{g^2}{2}}\right\},
\label{bose_lag_inst}
\end{eqnarray}
instead of (\ref{bose_lag}). This ambiguity arises from the extra
freedom due to the existence of two species of fermions, and does not exist in the
original massive Thirring model.  It is, however, easily seen that the 
Lagrangian above does not reproduce the expansion terms of
(\ref{Z_E}) for higher orders than $O({m''}^4)$ unless
$\alpha=1/2$ (see the Appendix). The boson model (\ref{bose_lag_orig}),
therefore, is a unique one which corresponds to the fermion model of
(\ref{lagrangian1}). It would be interesting to ask to what order one
should calculate to fix the corresponding boson model for the case of an 
$n$-species model.

Next we would like to refer to operator relations between fermion and 
boson fields. As usual, by comparing (\ref{bose_lag_orig}) with (\ref{lagrangian1}) we may put
\begin{eqnarray}
 \overline\psi_a\ooalign{\hfil/\hfil\crcr$\partial$}\psi_a &\Longleftrightarrow & \frac{g^2}{4\pi}(\partial\phi_a)^2,\\
\overline\psi_a\gamma_\mu\psi_a &\Longleftrightarrow & \epsilon_{\mu\nu}\partial^\nu\phi_a,
\end{eqnarray}
and the same for $b$ fields. These are the same as the original
one-species case for the massive Thirring and Sine-Gordon models. We
may also put
\begin{equation}
 m(\overline\psi_a\psi_a + \overline\psi_b\psi_b)\Longleftrightarrow {m'}^2\cos\left\{\sqrt{3g^2/2+g\pi}\phi_a'\right\}\cos\left\{\sqrt{3g^2/2-g\pi}\phi_b'\right\}.\label{mass_corres}
\end{equation}
It is difficult to separate the right-hand side of the above expression
into terms each of which contains only $\phi_a$ or $\phi_b$. We cannot,
therefore, make correspondences between the fermion and boson fields for each
species. It is hard to expect a correspondence like
(\ref{mass_corres}) from a naive extension of the original single-species
case. 

We have stated nothing about value of $g$ so far, except its sign. It seems 
that for $\pi>g>0$, $\phi_b'$ becomes a ghost field, because the
coefficient of its kinetic term in the Lagrangian (\ref{bose_lag}) is
negative for such values. From the viewpoint of the path integral
formulation, we see that the quantity $\int d\phi_b' \exp\{-S(m=0)\}$ is
ill-defined even if we are in  Euclidean space-time. To investigate a two-body
composite system, use of the Bethe-Salpeter (BS) equation is often
made. It is known that there is an abnormal solution for the BS equation
for  most cases. We conjecture here that the $\phi_b'$-field would
correspond to such an abnormal solution for the particle--anti-particle
composite state.
 
We find from (\ref{bose_lag}) that $\phi_b'$ becomes a free field for
$g=2\pi/3$, though it is in the ghost region mentioned above, while $\phi_a'$
becomes a Sine-Gordon field. Therefore, for such a specific value of $g$, our
system is equivalent to an ordinary Sine-Gordon  model with a single
species. A similar fact is seen for some $SU(2)$ extension of the
model. As mentioned above, if we take $(\sum_{i=a,b}j_i)^2$, which is
one of the diagonal terms of $SU(2)$, instead of $j_{a\mu}j_b^{\mu}$ in
(\ref{lagrangian1}), we only need a single auxiliary bose field, unlike
(\ref{lagrangian2}). The value of $g$, however, is not concerned with 
 this case. 

If we rescale the boson fields as
\begin{eqnarray}
 \phi_a'\rightarrow \phi_a'' &=& \sqrt{g^2/2\pi +g/2}\phi_a',\\
\phi_b'\rightarrow \phi_b'' &=& \sqrt{g^2/2\pi -g/2}\phi_b',
\end{eqnarray}
then we have from (\ref{bose_lag})
\begin{equation}
 {\cal L}= \frac{1}{2}(\partial\phi_a'')^2 + \frac{1}{2}(\partial\phi_b'')^2 +
\frac{2{m'}^2}{\pi}\cos\left\{\sqrt{\frac{3\pi g+2\pi^2}{g+\pi}}\phi_a''\right\}\cos\left\{\sqrt{\frac{3\pi g-2\pi^2}{g-\pi}}\phi_b''\right\}.
\end{equation}
For $g= 0$ we have
\begin{equation}
 {\cal L}= \frac{1}{2}(\partial\Phi_+)^2+\frac{1}{2}(\partial\Phi_-)^2 + 
\frac{{m'}^2}{\pi}\left\{\cos\sqrt{4\pi}\Phi_++\cos\sqrt{4\pi}\Phi_-\right\},
\end{equation}
where we have set
\begin{equation}
 \Phi_{\pm}=\frac{\sqrt{2}}{2}\left(\phi_a''\pm \phi_b''\right).
\end{equation}
This is the well-known boson model which corresponds to the free
fermion model.

\section*{Acknowledgments}
The authors would like to thank Professor T. Kugo and Professor S. Itoi for
useful comments. They also thank Dr. Shijong Ryang for informing them of
Ref.~\citen{bosonization}. 

\setcounter{equation}{0}
\appendix
\section{The $n$-th Order Terms of the Perturbative Expansion}
We now prove that the generating functional $Z_E$ of (\ref{Z_E}) is
equal to that of (\ref{bose_lag}) order by order in  the expansion with
respect to $m$ or $m'$. The $n$-th order term of $Z_E$ is given as
\begin{equation}
 Z_E^{(n)}=\frac{m^n}{n!}\int\prod dxdy\left\langle\left\{\overline\psi_ae^{ig\gamma_5\phi_a}\psi_a + \overline\psi_be^{ig\gamma_5\phi_b}\psi_b\right\}^n\right\rangle.
\end{equation} 
As is mentioned in the text, the odd order terms of the above expansion
vanish. Then we set $n=2k$ to obtain
\begin{eqnarray}
 Z_E^{(2k)}&=&\frac{m^{2k}}{(2k)!}\sum_{r=0}^k\frac{(2k)!}{(2r)!(2k-2r)!}\nonumber\\
&&\times\left\langle
\left\{\int dx\left(e^{ig\phi_a}\overline\psi_a\frac{1+\gamma_5}{2}\psi_a+e^{-ig\phi_a}\overline\psi_a\frac{1-\gamma_5}{2}\psi_a\right)\right\}^{2r}\right.\nonumber\\
&&\times\left.\left\{\int dx(a\rightarrow b)\right\}^{2k-2r}\right\rangle.
\end{eqnarray}
Due to the super selection rule for massless boson fields, the number of
$+\phi_a$  $(+\phi_b)$ must be equal to that of $-\phi_a$ $(-\phi_b)$ in
exponentials of the above expression, and thus we have
\begin{eqnarray}
 Z_E^{(2k)}&=&m^{2k}\sum_{r=0}^k\frac{1}{(r!)^2\{(k-r)!\}^2}\nonumber\\
&&\times\left\langle\left\{\int dxdye^{ig(\phi_a(x)-\phi_a(y))}\overline\psi_a(x)\frac{1+\gamma_5}{2}\psi_a(x)\overline\psi_a(y)\frac{1-\gamma_5}{2}\psi_a(y)\right\}^r\right. \nonumber\\
&&\times\left. \left\{\int dxdy e^{ig(\phi_b(x)-\phi_b(y))}\overline\psi_b\frac{1+\gamma_5}{2}\psi_b(x)\overline\psi_b\frac{1-\gamma_5}{2}\psi_b(y)\right\}^{k-r}\right\rangle\nonumber\\
&=&m^{2k}\sum_{r=0}^k\frac{1}{(r!)^2\{(k-r)!\}^2}\int\prod_{i=1}^k dx_idy_i\nonumber\\
&&\times\left\langle\prod_{i=1}^r\overline\psi_a(x_i)\frac{1+\gamma_5}{2}\psi_a(x_i)
\overline\psi_a(y_i)\frac{1-\gamma_5}{2}\psi_a(y_i)\right\rangle\nonumber\\
&&\times\left\langle\prod_{j=r+1}^k \overline\psi_b(x_j)\frac{1+\gamma_5}{2}\psi_b(x_j)\overline\psi_b(y_j)\frac{1-\gamma_5}{2}\psi_b(y_j)\right\rangle\nonumber\\
&&\times\left\langle\exp ig\left[\sum_{i=1}^r\left\{\phi_a(x_i)-\phi_a(y_i)\right\}+
\sum_{j=r+1}^k\left\{\phi_b(x_j)-\phi_b(y_j)\right\}\right]\right\rangle\nonumber
\label{Z_E(2k)}
\end{eqnarray}
\begin{eqnarray}
 &=&\left(\frac{m}{2\pi}\right)^{2k}\sum_{r=0}^{k}\frac{1}{(r!)^2\{(k-r)!\}^2}
\int\prod_{i=1}^k dx_idy_i \nonumber\\
&&\times\frac{\prod_{i>j=1}^{r}(x_i-x_j)^2(y_i-y_j)^2}{\prod_{i,j=1}^{r}(x_i-y_j)^2}\cdot\frac{\prod_{i>j= r+1}^k(x_i-x_j)^2(y_i-y_j)^2}{\prod_{i,j=r+1}^{k}(x_i-y_j)^2}\nonumber\\
&&\times\left\langle \exp ig\left[\sum_{i=1}^r\left\{\phi_a(x_i)-\phi_a(y_i)\right\}+\sum_{j=r+1}^k\left\{\phi_b(x_j)-\phi_b(y_j)\right\}\right]\right\rangle .\label{Z_E_final}
\end{eqnarray}
On the other hand, from the boson model (\ref{bose_lag}) we have
\begin{eqnarray}
Z_{E(B)}^{(n)}&=&\frac{{m'}^{2n}}{n!}\int\prod_i^n dx_i\left\langle \prod_i^n\cos\beta_+\phi_a'(x_i)\cos\beta_-\phi_b'(x_i)\right\rangle\nonumber\\
&=&\frac{1}{n!}\left(\frac{m'}{2}\right)^{2n} \int\prod_i^n dx_i\left\langle\prod_i^n\left\{e^{i\beta_+\phi_a'(x)}+e^{-i\beta_+\phi_a'(x_i)}\right\}\right\rangle\nonumber\\
&&\times \left\langle\prod_i^n\left\{e^{i\beta_-\phi_b'(x_i)}+e^{-i\beta_-\phi_b'(x_i)}\right\}\right\rangle ,\label{Z_EB}
\end{eqnarray}
where we have put
\begin{equation}
 \beta_{\pm}=\sqrt{3g^2/2\pm g\pi}.
\end{equation}
We can set $n=2k$, because only terms in the above expression 
containing the same numbers of $+\phi_a'$ and $-\phi_a'$ in
exponential do not vanish. We have
\begin{equation}
 \left\langle\prod_i^{2k}\left\{e^{i\beta_+\phi_a'(x_i)}+e^{-i\beta_+\phi_a'(x_i)}\right\}\right\rangle 
= \frac{(2k)!}{k!k!}\left\langle\exp i\beta_+\sum_i^k\{\phi_a'(x_i)-\phi_a'(y_i)\}\right\rangle,
\end{equation} 
where we have rearranged  arguments $\{x_1,x_2,\cdots ,x_{2k}\}\rightarrow
\{x_1,x_2,\cdots ,x_k,y_1,\cdots ,y_k\}$ so that the $x_i$ and $y_i$ are
arguments of the
$\{+\phi_a'\}$ and $\{-\phi_a'\}$, respectively. By this rearrangement, the last factor of (\ref{Z_EB})
becomes
\begin{eqnarray*}
 \sum_{r=0}^k\left(\frac{k!}{r!(k-r)!}\right)^2 \left\langle\exp i\beta_-\left[\sum_{i=1}^r\{\phi_b'(x_i)-\phi_b'(y_i)\}-\sum_{i=r+1}^k\{\phi_b'(x_i)-\phi_b'(y_i)\}\right]\right\rangle.
\end{eqnarray*}
Then we have
\begin{eqnarray}
 Z_{E(B)}^{(2k)}&=&(m')^{4k}\sum_{r=0}^k\left(\frac{1}{r!(k-r)!}\right)^2\int\prod dxdy
\left\langle \exp i\beta_+\sum_i^k\left\{\phi_a'(x_i)-\phi_a'(y_i)\right\}\right\rangle\nonumber\\
&\times&\left\langle \exp i\beta_-\left[\sum_{i=1}^r\{\phi_b'(x_i)-\phi_b'(y_i)\}-
\sum_{i=r+1}^k\{\phi_b'(x_i)-\phi_b(y_i)\}\right]\right\rangle.
\end{eqnarray}
Making use of the identities
\begin{equation}
 \beta_+^2\Delta_a'(x-y) + \beta_-^2\Delta_b'(x-y) = g^2\langle\phi_a(x)\phi_a(y)\rangle-\ln (x-y)^2\mu^2,\label{beta_plus}
\end{equation}
and
\begin{equation}
 \beta_+^2\Delta_a'(x-y) - \beta_-^2\Delta_b'(x-y) = g^2\langle\phi_a(x)\phi_b(y)\rangle,\label{beta_minus}
\end{equation}
we obtain
\begin{eqnarray}
 Z_{E(B)}^{(2k)}&=& \left(\frac{m}{2\pi}\right)^{2k}\sum_{r=0}^k\left(\frac{1}{r!(k-r)!}\right)^2\int\prod dxdy\nonumber\\
&&\times\frac{\prod_{i>j=1}^r(x_i-x_j)^2(y_i-y_j)^2\prod_{i>j=r+1}^{k}(x_i-x_j)^2
(y_i-y_j)^2}{\prod_{i,j=1}^r (x_i-y_i)^2\prod_{i,j= r+1}^k(x_i-y_i)^2}
\nonumber\\
&&\times \exp g^2\left[\sum_{i>j=1}^r\left\{\Delta_a(x_i-x_i) + \Delta_a (y_i-y_j)\right\}\right.\nonumber\\
&& +\sum_{i>j=r+1}^{k}\left\{\Delta_a(x_i-x_j)+\Delta_a(y_i-y_j)\right\}\nonumber\\
&&-\sum_{i,j=1}^r\Delta_a(x_i-y_j)-\sum_{i,j= r+1}^k\Delta_a(x_i-y_j)\nonumber\\
&&+\left.\sum_{i=1}^r\sum_{j=r+1}^k\left\{\Delta_{ab}(x_i-x_j)+\Delta_{ab}(y_i-y_j)
-\Delta_{ab}(x_i-y_j)-\Delta_{ab}(x_j-y_i)\right\}\right].\nonumber\\
{}\label{Z_EB_final}
\end{eqnarray}
where we put
\begin{eqnarray}
&\Delta_a(x-y)&\equiv\langle\phi_a(x)\phi_a(y)\rangle -\langle\phi_a(x)^2\rangle =\langle\phi_b(x)\phi_b(y)\rangle-\langle\phi_b(x)^2\rangle,\ \ \ \ \ \ \ \ \\
&\Delta_{ab}(x-y)&\equiv\langle\phi_a(x)\phi_b(y)\rangle-\langle\phi_a(x)\phi_b(x)\rangle.
\end{eqnarray}
Then we find Eq.~(\ref{Z_EB_final}) is equal to (\ref{Z_E(2k)}), i.e. $Z_E^{(2k)}=Z_{E(B)}^{(2k)}$.

In the above argument it is essential that Eqs.~(\ref{beta_plus}) and
(\ref{beta_minus}) hold. If we take (\ref{bose_lag_inst}) as the
corresponding boson model,  then putting
\begin{eqnarray}
& \eta_{+}&=\sqrt{2\alpha(g^2+g\pi)+g^2/2},\\
& \eta_{-}&=\sqrt{2(1-\alpha)(g^2-g\pi)+g^2/2}
\end{eqnarray} 
instead of $\beta_{+}$ and $\beta_{-}$ respectively,  we have
\begin{equation}
\eta_+^2\Delta_a'(x-y)+\eta_-^2\Delta_b'(x-y) =
 g^2\langle\phi_a(x)\phi_a(y)\rangle - \ln(x-y)^2\mu^2,\label{eta_plus}
\end{equation}
\begin{equation}
\eta_+^2\Delta_a'(x-y) -\eta_-^2\Delta_b'(x-y) = g^2\langle\phi_a(x)\phi_b(y)\rangle - (2\alpha-1)\ln (x-y)^2\mu^2,\label{eta_minus}
\end{equation}
instead of (\ref{beta_plus}) and (\ref{beta_minus}). Due to the last
term of (\ref{eta_minus}), $Z_E^{(2k)}$ is not equal to $Z_{E(B)}^{(2k)}$
unless $\alpha=1/2$.

\end{document}